# Quantum frequency comb with pump-selectable bin pairing and extraction-aware loading in a lithium niobate microresonator

Mohamad Reza Nurrahman[1, †], Hyeon Hwang[1, †, §], Nuri Han[1], Guhwan Kim[2], Hong-Seok Kim[2], Kiwon Moon[2], Jung Jin Ju[2], Hansuek Lee[1], and Min-Kyo Seo[1,*]

[1]Department of Physics, Korea Advanced Institute of Science and Technology, Daejeon, 34141, Republic of Korea

[2]Quantum Optics Research Section, Electronics and Telecommunications Research Institute, Daejeon, 34129, Republic of Korea

[†]These authors contributed equally to this work.

[§]hyeon_hwang@kaist.ac.kr

[*]minkyo_seo@kaist.ac.kr

## Abstract

Integrated quantum photonics requires bright, high-fidelity photon-pair sources capable of spectral multiplexing, correlation control, and circuit-compatible extraction. Cavity-enhanced spontaneous parametric down-conversion (SPDC) increases pair generation, but triply resonant operation imposes stringent pump-signal-idler spectral-alignment constraints. Moreover, the trade-off between intrinsic generation and coincidence-to-accidental ratio (CAR) does not capture the usable output flux, which depends on photon extraction. Here, we demonstrate a single-pass-pumped, resonator-enhanced quantum frequency comb (QFC) source based on a periodically poled lithium niobate photonic-crystal Fabry-Pérot microresonator. The device yields intrinsic and loaded brightnesses of 69.9 and 1.88 MHz/μW, respectively, and a maximum CAR of 16,000. Frequency-resolved measurements reveal 461 cavity-defined bins spanning 1495-1570 nm, and loaded spectral brightness approaching $2.22\times10^9$ pairs/(s·mW·nm). Heralded autocorrelation measurements yield a second-order autocorrelation $g_H^{(2)}(0)$ as low as $4.0\times10^{-4}$ for a selected high-CAR cavity-defined bin pair. In particular, tuning the single-pass pump deterministically selects correlated frequency-bin pairings within the fixed QFC grid while preserving brightness and pairwise coincidence rates. We further separate intrinsic generation from output photon-pair flux, revealing the loaded-brightness-CAR relation. Together, pump-selectable bin pairing and extraction-aware loading point to tailored SPDC QFCs as chip-integrated nonclassical light resources for multichannel quantum-state transport and information processing.

## Introduction

Quantum information science and technology increasingly demands hardware capable of generating, distributing, and processing nonclassical states with high fidelity and scalability,[1,2] enabling secure communication,[3,4] enhanced information processing,[5,6] high-precision sensing,[7,8,9] and quantum computation.[10,11,12,13] Photons are particularly attractive carriers of quantum information owing to their long coherence times, high-speed transmission, and compatibility with existing optical communication infrastructure.[14,15,16] Nonclassical states of light, including entangled photon pairs and squeezed states,[17,18,19] have therefore become central resources for optical quantum communication, networks, and information processing.[20,21,22,23] As photonic quantum systems move toward programmable and multiplexed architectures, however, the key requirement is no longer only to generate nonclassical light, but also to control its spectral, temporal, modal, and correlation structure.[24,25,26] Extending correlated photon-pair generation across multiple frequency channels further increases usable optical bandwidth and enables wavelength- and frequency-bin-multiplexed quantum functionalities.[27,28,29,30] On-chip realization of such sources requires high brightness, low noise, well-defined spectral modes, efficient extraction into usable photonic-circuit modes, and control over the generated frequency-bin pairing.[31,32,33,34]

Integrated nonlinear photonics provides a natural route to compact and scalable photon-pair sources.[29-40] $\chi^{(2)}$-based spontaneous parametric down-conversion (SPDC) offers high nonlinear efficiency, quasi-phase-matched wavelength engineering, and low nonlinear noise, compared with $\chi^{(3)}$-based spontaneous four-wave mixing.[31,41,42] The large spectral separation between the pump and generated signal-idler photons facilitates efficient pump rejection.[43,44] Thin-film lithium niobate (TFLN) has been especially attractive because of its strong second-order nonlinearity, low optical loss, broadband transparency, electro-optic functionality, and mature periodic-poling technology.[45,46,47] Optical cavities enhance SPDC by increasing the density of optical states for the signal and idler photons and reshaping broadband emission into discrete frequency-bin pairs.[48,49] Recent periodically poled lithium-niobate (PPLN) and lithium-tantalate (PPLT) traveling-wave resonators, including microring and racetrack geometries, have demonstrated high-brightness, high-quality multi-wavelength photon-pair generation.[31,50,51,52,53] In triply resonant implementations, however, a short-wavelength pump resonance must be aligned with telecom signal and idler resonances that experience distinct dispersion, imposing stringent spectral-matching constraints and limiting control of frequency-bin pairing. Moreover, prior benchmarks have mostly emphasized intrinsic generation rate, whereas usable photon-pair flux depends on the loaded generation rate at the output.

Photonic-crystal Fabry–Pérot (PhC-FP) cavities provide a distinct geometry for addressing these challenges. Recent works have realized high-performance TFLN standing-wave PhC-FP resonators with intrinsic quality (Q) factors exceeding $10^6$,[54,55,56,57] motivating their use for cavity-enhanced quantum light generation. In our approach, X-cut PPLN is integrated with a one-dimensional PhC-FP microresonator whose photonic bandgap selectively confines the telecom-band signal and idler photons, while the short-wavelength pump propagates in a single-pass configuration. This single-pass-pumped, signal-idler-resonant architecture relaxes the pump-signal-idler resonance-matching requirement, imposed by triply resonant cavities, and the pump frequency can further serve as an independent knob for selecting bin pairs within the fixed signal-idler grid. Within the bandgap, the PhC mirrors provide a fundamental-mode-selective resonance ladder with relatively uniform high-Q modes and a predictable loading profile, defining the primary operating window for a reliable frequency-bin grid for integrated quantum-light applications.

Here, we demonstrate bright quantum frequency comb (QFC) generation and photon-pair extraction in a single-pass-pumped PPLN PhC-FP microresonator. Mode-integrated SPDC measurements yield intrinsic and loaded brightnesses of 69.9 and 1.88 MHz/μW, respectively, together with a maximum coincidence-to-accidental ratio (CAR) of 16,000. Frequency-resolved measurements identify 461 cavity-defined frequency bins spanning 1495-1570 nm, with strong pairwise correlations and loaded spectral brightness approaching $2.22 \times 10^9$ pairs $s^{-1}$ $mW^{-1}$ $nm^{-1}$. Selected-bin measurements of the heralded second-order autocorrelation yield $g_H^{(2)}(0)$ down to $(4.0 \pm 1.1) \times 10^{-4}$. Beyond generating a dense and bright comb, we demonstrate pump-selectable frequency-bin pairing within a fixed spectral grid defined by the signal-idler resonance ladder. By separating the intrinsic photon-pair generation rate from the loaded pair flux extracted to the output, we also reveal that the usable brightness-CAR relation is governed not only by cavity-enhanced nonlinear generation, but also by the cavity-loading condition. These achievements establish the single-pass-pumped PhC-FP architecture as a compact source platform for pump-selectable, extraction-aware SPDC QFCs whose correlation structures can be tailored by pump shaping and further processed by downstream photonic circuits.

## Results

### PPLN PhC-FP microresonator platform for quantum frequency comb generation

Efficient QFC generation in a cavity requires simultaneous control of three spectral functions: the quasi-phase-matched $\chi^{(2)}$ gain, the cavity-enhanced density of signal-idler modes, and the extraction efficiency that determines the usable output rate. For programmable frequency-bin resources, the source should also provide a way to select the correlated bin-pairing rule without perturbing the signal-idler resonance grid. We implement this control using a single-pass-pumped PPLN PhC-FP microresonator, schematically shown in Fig. 1a. A continuous-wave pump at the second-harmonic wavelength is launched into a 3-mm-long PPLN waveguide through the wavelength-division multiplexer (WDM) based on adiabatic directional coupler, where it drives SPDC processes in a single-pass configuration. In contrast, the down-converted telecom photons are resonantly confined by an asymmetric PhC-FP cavity formed by a high-reflectivity input-side mirror and a partial reflector at the output side. The overlap between the broadband quasi-phase-matched SPDC gain and the longitudinal modes of the PhC-FP cavity defines a dense set of frequency-correlated signal-idler mode pairs, while the partial reflector loads the cavity and extracts the generated photons into a single output port.

A key feature of our platform is that the PhC mirrors do more than provide high reflectivity. Their bandgap reflectivity, Bloch-mode phase response, and mirror penetration depth vary across the comb spectrum, producing a frequency-dependent loaded-Q landscape. As a result, different frequency-bin pairs experience different combinations of cavity enhancement and extraction efficiency. This provides a direct route to distinguish the intrinsic pair-generation rate, $PGR_{\mathrm{int}}$, from the loaded pair-generation rate, $PGR_{\mathrm{load}}$, and establishes a basis for coupling-engineered, extraction-aware optimization of usable brightness and CAR.

**Fig. 1: Integrated PPLN PhC-FP microresonator for SPDC quantum frequency comb generation**

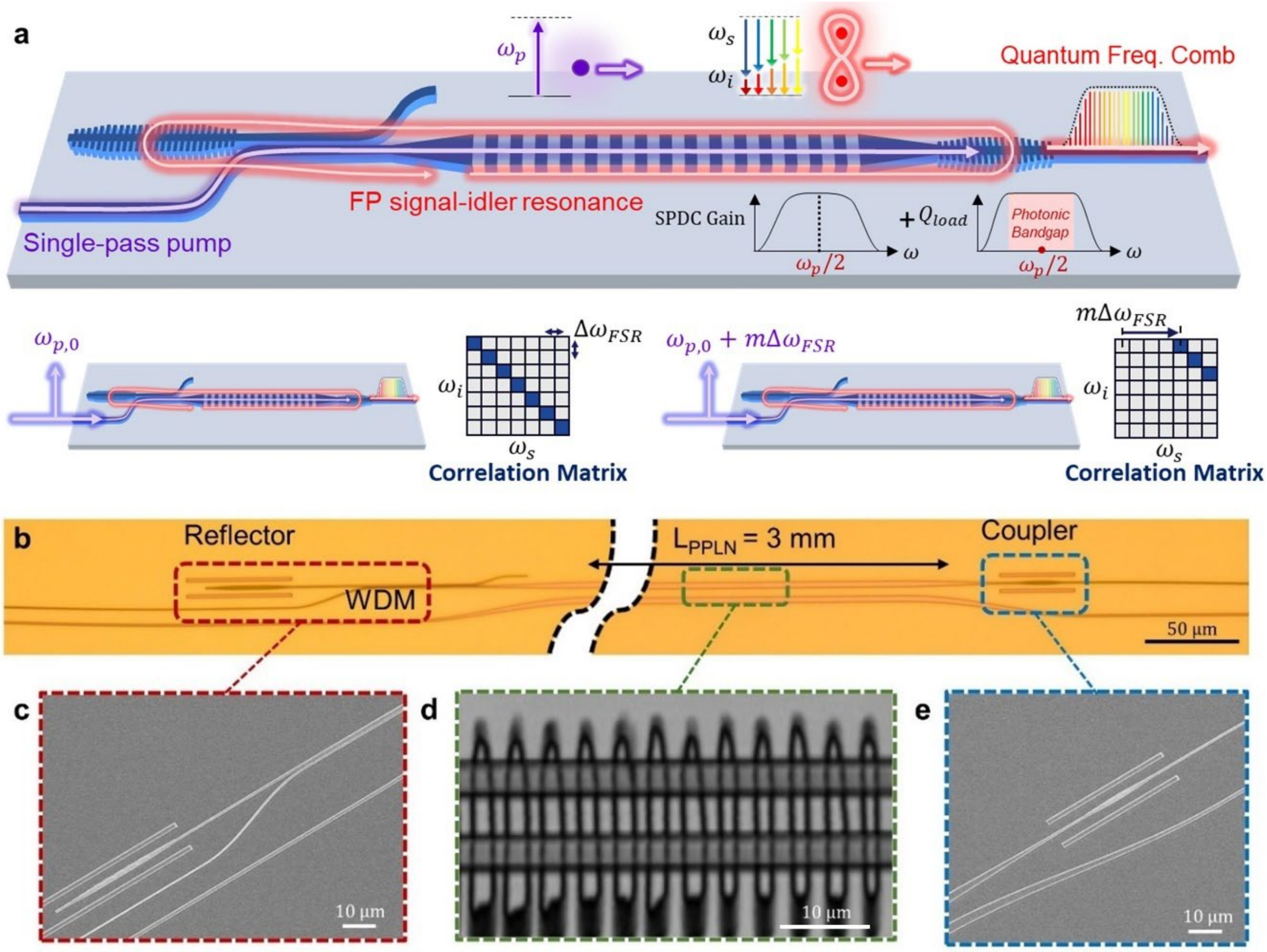

**a.** Schematic of cavity-enhanced SPDC-based quantum frequency comb generation and pump-selectable frequency bin pairing in an integrated PPLN PhC-FP microresonator. A visible pump drives SPDC in the PPLN section, while the generated telecom-band photon pairs are resonantly enhanced by the PhC-FP cavity and extracted through the output port. By tuning the visible pump frequency by amounts corresponding to integer multiples of the free spectral range (FSR) in the signal-idler band, the frequency-bin pairing can be selectively controlled. **b.** Optical microscope image of the fabricated device, showing the WDM and high-reflectivity PhC mirror, the 3-mm-long PPLN section (green dashed box), and the output-side partial PhC reflector (blue dashed box). **c.** SEM image of the high-reflective PhC/WDM region. **d.** confocal SHG microscopy image of the PPLN region. **e.** SEM image of the partial PhC reflector used for output coupling.

The fabricated device is shown in Fig. 1b. The PPLN section is integrated between the input-side WDM/high-reflectivity PhC mirror and the output-side partial PhC reflector, forming an asymmetric standing-wave cavity for the telecom-band SPDC photons. Scanning electron microscope (SEM) images of the WDM/high-reflective PhC mirror and the partially reflective output PhC mirror for extracting the down-converted photons are shown in Fig. 1c and 1e, respectively. The periodic domain inversion in the nonlinear optical section is verified by second-harmonic-generation (SHG) confocal microscopy, as shown in Fig. 1d. A reference PPLN waveguide fabricated next to the PhC-FP microresonator is used to independently characterize the quasi-phase-matching spectrum of SHG and to align the SPDC pump wavelength to the maximum nonlinear gain.

## Mode-integrated SPDC generation and reciprocal SHG calibration

We first quantify the mode-integrated photon-pair generation rate (PGR) of the PPLN PhC-FP microresonator device before resolving individual frequency-bin pairs. The device is pumped at 766 nm and operated at 60 °C, where the detected telecom-band SPDC photon count is optimized. The measurement setup is schematically shown in Fig. 2a. The pump laser wavelength is actively stabilized using a Fizeau wavemeter with PID feedback, suppressing the frequency fluctuation to within ±15 MHz around the target wavelength. At the device output, the residual pump is separated from the SPDC telecom photons using an optical fiber WDM and further rejected by three cascaded long-pass filters, each providing approximately 50 dB attenuation for wavelengths below 1050 nm. The SPDC photons are then split by a 50:50 beam splitter and directed to two separate channels of superconducting nanowire single-photon detectors (SNSPDs), while the detection events are recorded by a time tagger.

Figure 2b shows the measured single-photon count rates $S_1$ and $S_2$ in the two detection channels and the corresponding true coincidence count rate $C$ as a function of on-chip pump power. For the 50:50 beam-splitter measurement, the detected single-photon count rate in channel $i = 1, 2$ is $S_i =$ $2PGR_{\text{int}}\eta_{\text{ext}}\mu_i\eta_{D,i}$, where $PGR_{\text{int}}$ is the intrinsic PGR inside the cavity, $\eta_{\text{ext}}$ is the extraction efficiency of the SPDC photons through the output coupler, $\mu_i$ is the calibrated optical transmission from the bus waveguide immediately after cavity loading to the detector input in channel $i$, including the beam-splitter path and filtering losses, and $\eta_{D,i}$ is the SNSPD detection efficiency. The corresponding coincidence count rate is given by $S_c = 2PGR_{\text{int}}\eta_{\text{ext}}^2\mu_1\mu_2\eta_{D,1}\eta_{D,2}$. The factor of 2 in these expressions accounts for the two photons generated in each pair and the two possible beam-splitter output configurations contributing to coincidences.[51] From the measured $S_1$, $S_2$, and $S_C$, the intrinsic PGR is obtained directly as

$$PGR_{\text{int}} = \frac{S_1S_2}{2S_c} \tag{1}$$

This expression is independent of the optical propagation efficiencies, detector efficiencies, and extraction efficiency because these factors cancel in the ratio.

The loaded photon-pair generation rate, $PGR_{\text{load}}$, is defined as the pair flux extracted through the output coupler before off-chip propagation and detection losses. For the mode-integrated measurement,

$$PGR_{\text{load}} = \eta_{\text{ext}}^2 PGR_{\text{int}} = \frac{S_c}{2\mu_1\mu_2\eta_{D,1}\eta_{D,2}} \tag{2}$$

Unlike $PGR_{\text{int}}$, $PGR_{\text{load}}$ requires calibration of the optical transmission and detector efficiencies. In our present setup, the measured collection efficiencies from the device output facet to the SNSPDs are $\mu_1$ = 6.97% and $\mu_2$ = 7.29% (−11.57 dB and −11.37 dB), and the corresponding SNSPD detection efficiencies are $\eta_{D,1}$ = 84% and $\eta_{D,2}$ = 86%. In this mode-integrated measurement, $\eta_{\text{ext}}$ represents an effective average extraction efficiency. Frequency-bin-dependent extraction of the signal and idler photons, and its impact on the mode-resolved loaded PGR, are discussed in detail in the CAR-PGR analysis associated with Fig. 6.

Figure 2c plots both $PGR_{\text{int}}$ and $PGR_{\text{load}}$ together with the corresponding CAR. Linear fits to the PGRs as a function of the on-chip pump power $P_{\text{pump}}$ yield an intrinsic brightness of $B_{\text{int}}$ = 69.9 MHz/μW and a loaded brightness of $B_{\mathrm{load}}$ = 1.88 MHz/μW, respectively. Notably, the loaded brightness places the PPLN PhC-FP architecture among the brightest single-pass-pumped, resonator-enhanced SPDC sources reported to date.[58] Resonant enhancement of the visible pump can further increase SPDC rates, but simultaneous alignment of the pump, signal, and idler resonances requires stringent spectral control and complicates device-to-device integration. The CAR decreases with increasing pump power, as expected from multi-pair generation in SPDC.[31,38,51,52] The device maintains a CAR 15.2 at $P_{\mathrm{pump}}$ = 1.95 μW, while the CAR reaches 16,000 at $P_{\mathrm{pump}}$ = 1.23 nW. The corresponding coincidence histogram in Fig. 2d, acquired over 50 min, shows a pronounced coincidence peak with a background level of approximately one count per time bin.

The SPDC pump condition is independently calibrated using reciprocal SHG measurements. In the SHG setup shown in Fig. 2e, a telecom-band pump is injected from the output coupler side, and the generated SHG signal is collected from the opposite facet. A reference PPLN waveguide is first used to identify the quasi-phase-matching envelope and verify that the corresponding second-harmonic wavelength lies within the tuning range of the visible SPDC pump laser, 765-781 nm. As shown in Fig. 2f, the SHG peak shifts from 1528 nm at 25 °C to 1532 nm at 60 °C. The 1532-nm telecom pump produces a second-harmonic wavelength of 766 nm, matching the optimized SPDC pump wavelength used in the photon-pair measurements. The reference-waveguide measurement therefore fixes the device temperature and visible-pump wavelength for maximum nonlinear gain.

The PhC-FP cavity is characterized under the same temperature and wavelength conditions. The cavity SHG spectrum (black line in Fig. 2f) is consistent with that of the reference waveguide. At 60 °C, the cavity-enhanced SHG spectrum overlaps with an infrared cavity resonance near 1532 nm, as confirmed by the reflectance spectrum in Fig. 2g. This overlap verifies simultaneous alignment of the quasi-phase-matched $\chi^{(2)}$ gain and the telecom-band cavity mode. On resonance, the device generates up to 1.2 nW of second-harmonic power at an on-chip input pump power of 1.75 μW. A quadratic fit yields an SHG conversion efficiency of 36,240%/W (Fig. 2h), confirming efficient nonlinear interaction in the PPLN PhC-FP microresonator.

**Fig. 2: Total photon-pair brightness and SHG-calibrated nonlinear gain**

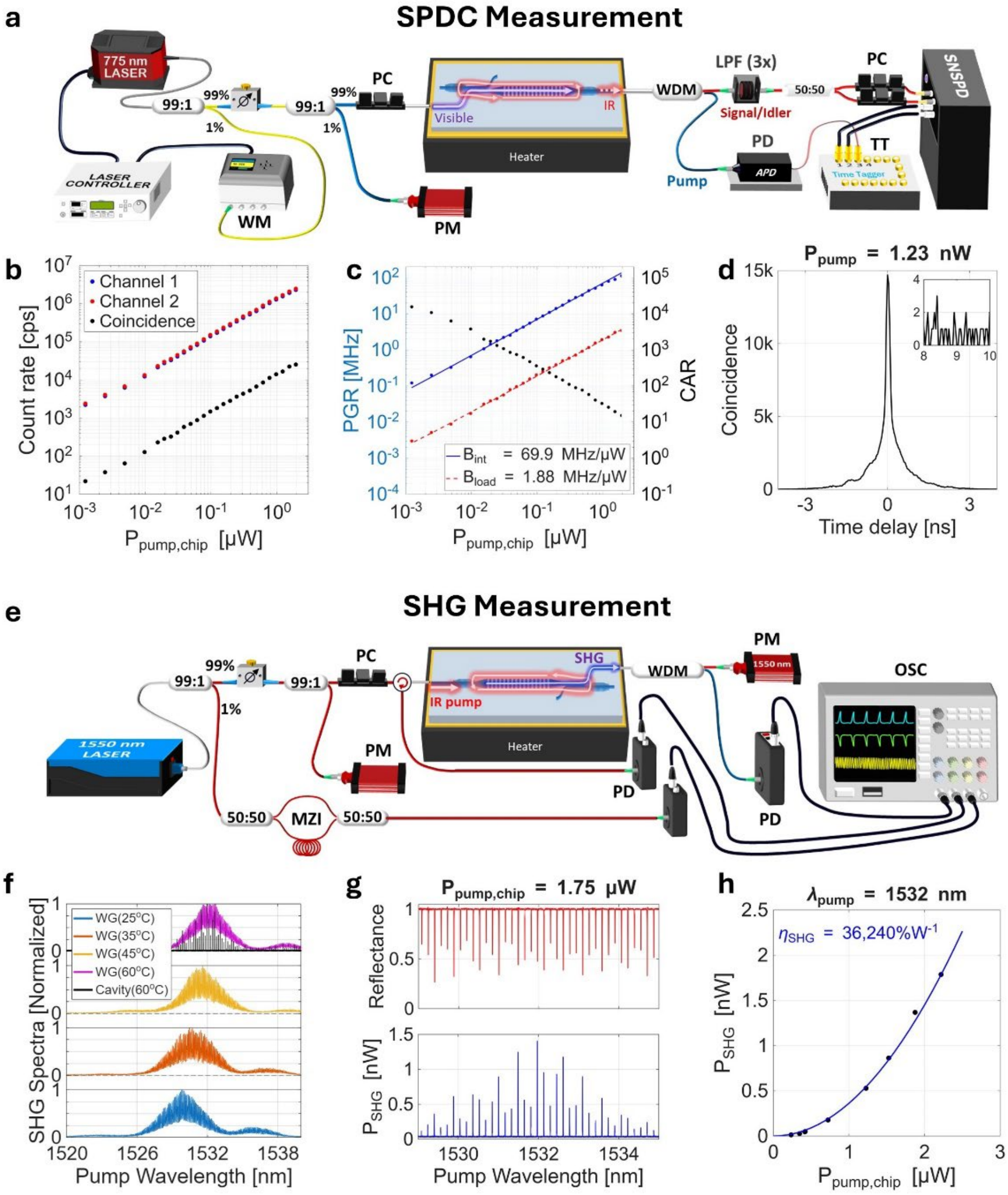

**a** Experimental setup for mode-integrated SPDC photon-pair generation measurements. **b** Measured single-photon count rates at channel 1 (blue dots) and channel 2 (red dots), and true coincidence rate (black dots) as a function of on-chip pump power. **c** Measured intrinsic PGR, loaded PGR, and CAR. **d** Coincidence histogram measured at an on-chip pump power of 1.23 nW. **e** Experimental setup for reciprocal SHG characterization. **f** Temperature-dependent SHG spectra of the reference waveguide (WG) and the PhC-FP cavity. **g** SHG power (blue) and reflectance (red) spectra of the PPLN PhC-FP microresonator measured at 60 °C. **h** On-chip SHG conversion efficiency as a function of telecom pump power. WM, wavemeter; PC, polarization controller; BS, beam splitter; PD, photodetector; PM, power meter; LPF, long-pass filter; SNSPD, superconducting nanowire single-photon detector; TT, time tagger; MZI, Mach–Zehnder interferometer; WDM, wavelength-division multiplexer; OSC, oscilloscope

## Dense quantum frequency comb with pairwise correlations

We next characterize the frequency-resolved structure of the SPDC QFC. Before measuring the photon-pair spectrum, the telecom-band cavity resonances are mapped by reflection spectroscopy from the output-coupler side, the same port used for SPDC measurements (Fig. 3a). The PhC-FP cavity supports a dense set of high-Q resonances across the photonic bandgap, with intrinsic Q factors $Q_i$ exceeding $10^6$. The coupling Q factor $Q_c$, extracted from fits to the reflectance spectra, varies with wavelength. The resonances are predominantly undercoupled inside the bandgap, where the PhC mirrors provide strong confinement. Toward and beyond the bandgap edge, reduced mirror reflectivity increases the output loading, driving the cavity through critical coupling toward the overcoupled regime. The Q-factor-derived extraction efficiency, $\eta_{\text{ext}} = Q_L/Q_c$, agrees with an independent estimate of photon extraction using the Klyshko method[59], validating the loading analysis used below. This landscape provides the resonant mode structure for QFC generation and underlies the mode-resolved CAR-$PGR_{\text{load}}$ relation.

The SPDC frequency comb is measured at an on-chip pump power of 200 μW. Two independent tunable bandpass filters (TBPFs), each with a bandwidth of 13 GHz, are placed in the two output arms of the 50:50 beam splitter before the SNSPDs. For the single-channel spectrum in Fig. 3b, one TBPF is scanned in steps of 1.23 GHz to resolve the cavity-defined frequency bins. The measured spectrum exhibits a broad comb envelope extending from 1495 to 1570 nm. The resonance peaks define 461 cavity-defined frequency bins across the measured spectral range, indexed from −230 to +230 relative to the degeneracy point. The bin spacing is approximately 21 GHz, set by the free spectral range (FSR) of the fundamental $TE_{00}$-resonance ladder ($\Delta\omega_{FSR}/2\pi$) that defines the comb grid.

To measure pairwise frequency correlations, the two TBPFs are tuned to conjugate frequency-bin pairs $[i, -i]$. Figure 3c shows the coincidence counts (dark bars) recorded for these correlated bin pairs across the full comb, together with the corresponding CAR (cyan dots) and resonance-linewidth-derived correlation time (red dots). The coincidence counts follow the cavity-enhanced SPDC envelope, confirming broadband generation of energy-conserving photon pairs. At the fixed pump power used for the spectral scan, the CAR varies strongly across the comb: the CAR is approximately 20.9 near the photonic bandgap center and increases toward the band edge, reaching values up to 5,200. This trend reflects the frequency-dependent cavity-loading landscape, which modifies the resonant enhancement, loaded linewidth, and extraction efficiency.

**Fig. 3: Cavity-defined comb grid and frequency-bin correlations**

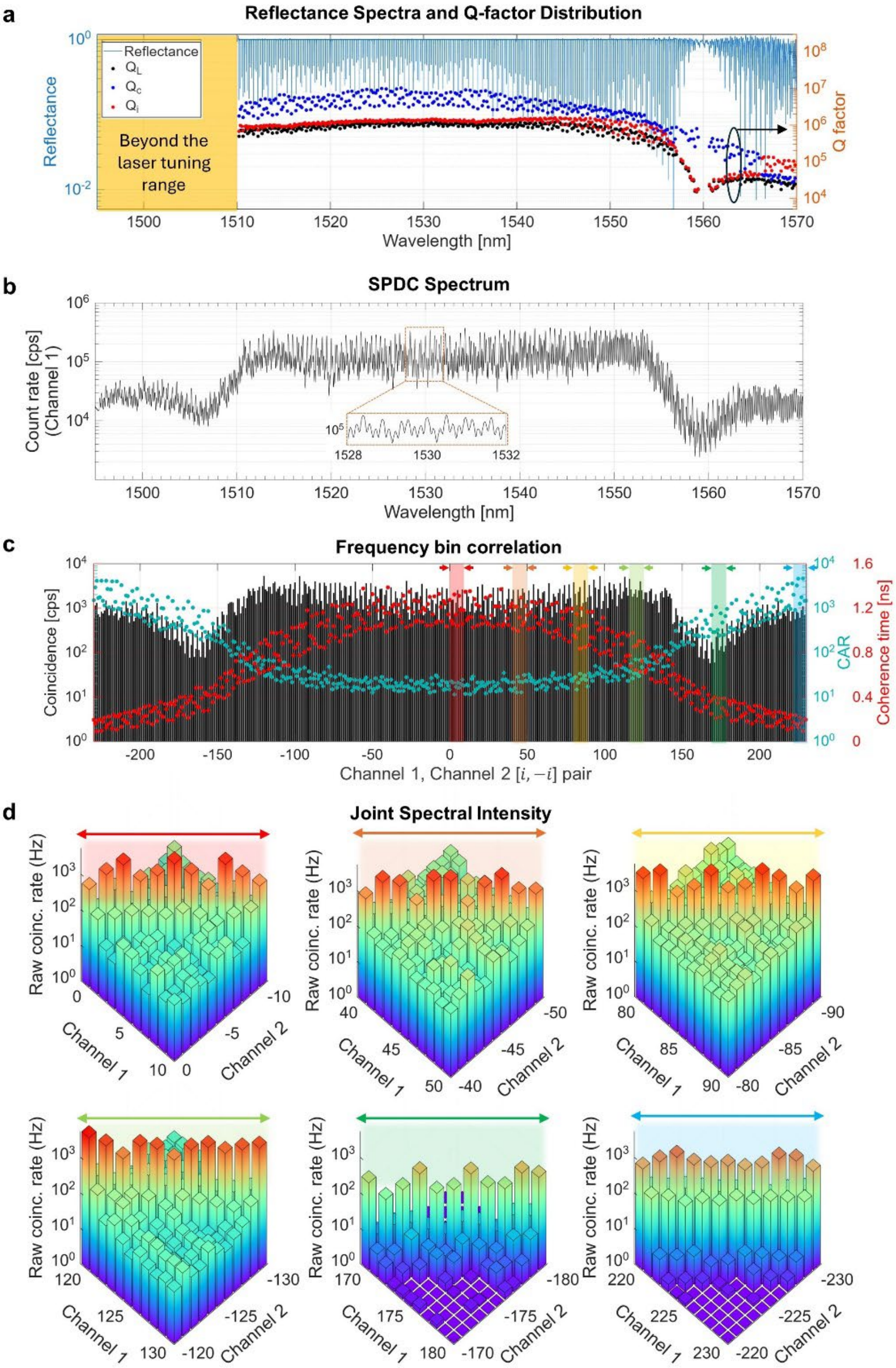

**a** Infrared reflectance spectrum (blue line) and extracted loaded, coupling, and intrinsic Q factors of the PPLN PhC-FP microresonator. **b** Single-channel SPDC spectrum of quantum frequency comb with a 13-GHz TBPF at an on-chip pump power of 200 μW. **c** Coincidence counts of conjugate frequency-bin pairs (black bars) and the corresponding CAR (cyan dots) and resonance-linewidth-derived coherence time (red dots). **d** Frequency-frequency correlation maps of 11 × 11 local frequency-bin windows in six representative spectral regions. Colored arrows indicate the spectral windows corresponding to the colored shaded regions in **c**. $Q_L$, loaded Q factor; $Q_i$, intrinsic Q factor; $Q_c$, coupling Q factor.

To further verify the pairwise frequency correlations, we measure frequency-frequency correlation maps in six representative spectral regions (Fig. 3d). In each region, the two TBPFs are independently scanned over an 11 × 11 local frequency-bin window, and the coincidence counts are recorded for all bin combinations. This two-dimensional test is important because high total coincidence counts alone do not establish channel selectivity in a dense frequency-bin comb. The largest coincidence counts occur at the energy-conserving conjugate-bin positions, while neighboring uncorrelated bin combinations remain close to the background level.

The contrast of the correlation maps follows the CAR distribution in Fig. 3c. In the band-edge windows, $i$ = 170–180 and $i$ = 220–230, where the CAR exceeds $10^3$, the correlation maps show strong suppression of uncorrelated-bin coincidences and pronounced high-contrast peaks. By contrast, the central-to-mid-bandgap windows spanning $i$ = 0–130 exhibit CAR values below $10^2$ and correspondingly lower contrast between correlated and uncorrelated frequency-bin combinations.

For the highest-index window, $i$ = 220–230, weak off-diagonal features are visible for neighboring bin combinations such as [$i$+1, −$i$] and [$i$−1, −$i$]. These features arise from the finite spectral selectivity of the measurement. In this region, the cavity resonance linewidth is ~9.6 GHz, comparable to the 13-GHz TBPF bandwidth and a substantial fraction of the 21-GHz FSR; the selected frequency-bin window partially overlaps with adjacent resonances, allowing neighboring-bin photons to contribute to the measured coincidence counts.

## Frequency-bin-resolved intrinsic and loaded brightness

Having established strong pairwise correlations across the SPDC QFC, we quantify the brightness of individual frequency-bin pairs. Six representative conjugate-bin pairs, [2, −2], [48, −48], [87, −87], [120, −120], [179, −179], and [228, −228], are selected from different spectral regions of the comb. For each pair, the two TBPFs are fixed to the corresponding signal and idler bins, and the single-photon count rates in both channels 1 and 2 and the coincidence counts are measured as a function of on-chip pump power.

Figure 4a shows the resulting $PGR_{\mathrm{int}}$, $PGR_{\mathrm{load}}$, and CAR for the selected frequency-bin pairs. Starting from the SPDC interaction Hamiltonian, the non-degenerate $PGR_{\mathrm{int}}^{(i)}$ for a single-pass pump configuration and resonant frequency-bin pair $[i, -i]$ is given by

$$PGR_{\mathrm{int}}^{(i)} = \frac{8\left|g_{i,-i}\right|^2}{\left(\kappa_{\mathrm{tot},i} + \kappa_{\mathrm{tot},-i}\right)} \left( \frac{L_{\mathrm{path}}}{v_{g,\mathrm{pump}}} \frac{P_{\mathrm{pump}}}{\hbar\omega_p} \right) \tag{3}$$

where $g_{i,-i}$ denotes the nonlinear coupling strength for the conjugate resonance pair, including the quasi-phase-matched $\chi^{(2)}$ overlap, and $\kappa_{\mathrm{tot},i}$ and $\kappa_{\mathrm{tot},-i}$ are the total energy decay rates of the signal and idler modes, $L_{\mathrm{path}}$ is the effective single-pass interaction length, $v_{g,\mathrm{pump}}$ is the group velocity at the pump wavelength, and $P_{\mathrm{pump}}/\hbar\omega_p$ is the incident pump photon flux, respectively. Equation (3) shows that, for slowly varying nonlinear overlap, $PGR_{\mathrm{int}}^{(i)}$ is enhanced by narrow signal and idler resonances and highlights the role of cavity enhancement.

The $PGR_{\mathrm{load}}$ is further determined by the extraction efficiencies of the signal and idler photons:

$$PGR_{\mathrm{load}}^{(i)} = \eta_{\mathrm{ext},i} \cdot \eta_{\mathrm{ext},-i} PGR_{\mathrm{int}}^{(i)} = \eta_{\mathrm{joint}}^{(i)} PGR_{\mathrm{int}}^{(i)} \tag{4}$$

The extraction efficiency of frequency bin $i$ is given by $\eta_{\mathrm{ext},i} = \frac{\kappa_{\mathrm{cpl},i}}{\kappa_{\mathrm{tot},i}} = \frac{Q_{L,i}}{Q_{\mathrm{c},i}}$, where $Q_{L,i}$ is the loaded Q factor and $Q_{\mathrm{c},i}$ is the coupling quality factor extracted from the reflectance analysis. $\eta_{\mathrm{joint}}^{(i)} = \eta_{\mathrm{ext},i} \cdot \eta_{\mathrm{ext},-i}$ is the joint extraction efficiency for the frequency-bin pair $[i, -i]$. Equation (4) predicts that the spectral trends of $PGR_{\mathrm{int}}$ and $PGR_{\mathrm{load}}$ can differ substantially, because the loaded rate includes the frequency-dependent extraction efficiencies of both photons. As a result, $PGR_{\mathrm{int}}$ is primarily governed by the nonlinear coupling and resonance linewidths, whereas $PGR_{\text{load}}$ is additionally engineered by the wavelength-dependent cavity-loading landscape.

This distinction is clearly observed in Fig. 4a. Within the photonic bandgap, $B_{\text{int}}$ decreases gradually from 124 kHz/μW for [2, −2] to 39.8 kHz/μW for [120, −120]. Outside the photonic bandgap, $B_{\text{int}}$ decreases to 5.38 kHz/μW for [179, −179] and 3.63 kHz/μW for [228, −228]. This trend follows Eq. (3): reduced mirror confinement toward and beyond the bandgap edge broadens the resonances and weakens the intrinsic cavity enhancement.

In contrast, $B_{\text{load}}$ follows a different spectral dependence. Although $B_{\text{int}}$ is largest near the bandgap center, the loaded-to-intrinsic brightness ratio, $B_{\text{load}}/B_{\text{int}}$, increases toward larger frequency-bin indices. For example, $B_{\text{load}}$ increases from 2.31 kHz/μW for [2,−2] to 11.1 kHz/μW for [87,−87], despite a decrease in $B_{\text{int}}$, over the same spectral range. Beyond the photonic bandgap edge, $B_{\text{int}}$ is substantially reduced, but $B_{\text{load}}/B_{\text{int}}$ remains enhanced because the cavity is more strongly loaded to the output channel. The different trends of $B_{\text{int}}$ and $B_{\text{load}}$ show that high internal cavity-enhanced generation and high usable output are governed by different aspects of the PhC-FP cavity response. The wavelength-dependent extraction efficiency, scaling as $Q_L/Q_c$, therefore directly translates into the observed changes in $B_{\text{load}}/B_{\text{int}}$.

To compare the source performance across frequency bins with different resonance linewidths, we examine the spectral brightness, the brightness normalized to the loaded resonance linewidth $\Delta\lambda$ of each selected signal-idler resonance pair. The intrinsic and loaded spectral brightness across the comb are shown in Fig. 4b. Square markers are obtained directly from the linear fits in Fig. 4a, whereas circular markers are estimated from the mode-resolved PGRs measured at an on-chip pump power of 200 μW. The good agreement between the two methods confirms the consistency of the PGR extraction and linewidth normalization.

$B_{\text{int}}/\Delta\lambda$ exceeds $4.37 \times 10^{10}$ pairs/(s·mW·nm) near the center of the photonic bandgap, where high-Q resonances provide the strongest nonlinear enhancement. The maximum $B_{\text{load}}/\Delta\lambda$ reaches $2.22 \times 10^{9}$ pairs/(s·mW·nm) near the photonic band edge, where increased output loading improves usable photon-pair extraction. Notably, the spectral-brightness map provides a guideline for designing cavity-enhanced SPDC QFCs: high intrinsic Q enhances pair generation, but the usable output flux is optimized only when resonant enhancement is balanced with photon extraction. By sampling multiple loading regimes in a single device, the present PhC–FP platform maps how resonant enhancement and photon extraction should be balanced for brighter SPDC QFCs.

**Fig. 4: Loading-dependent pair generation and spectral brightness**

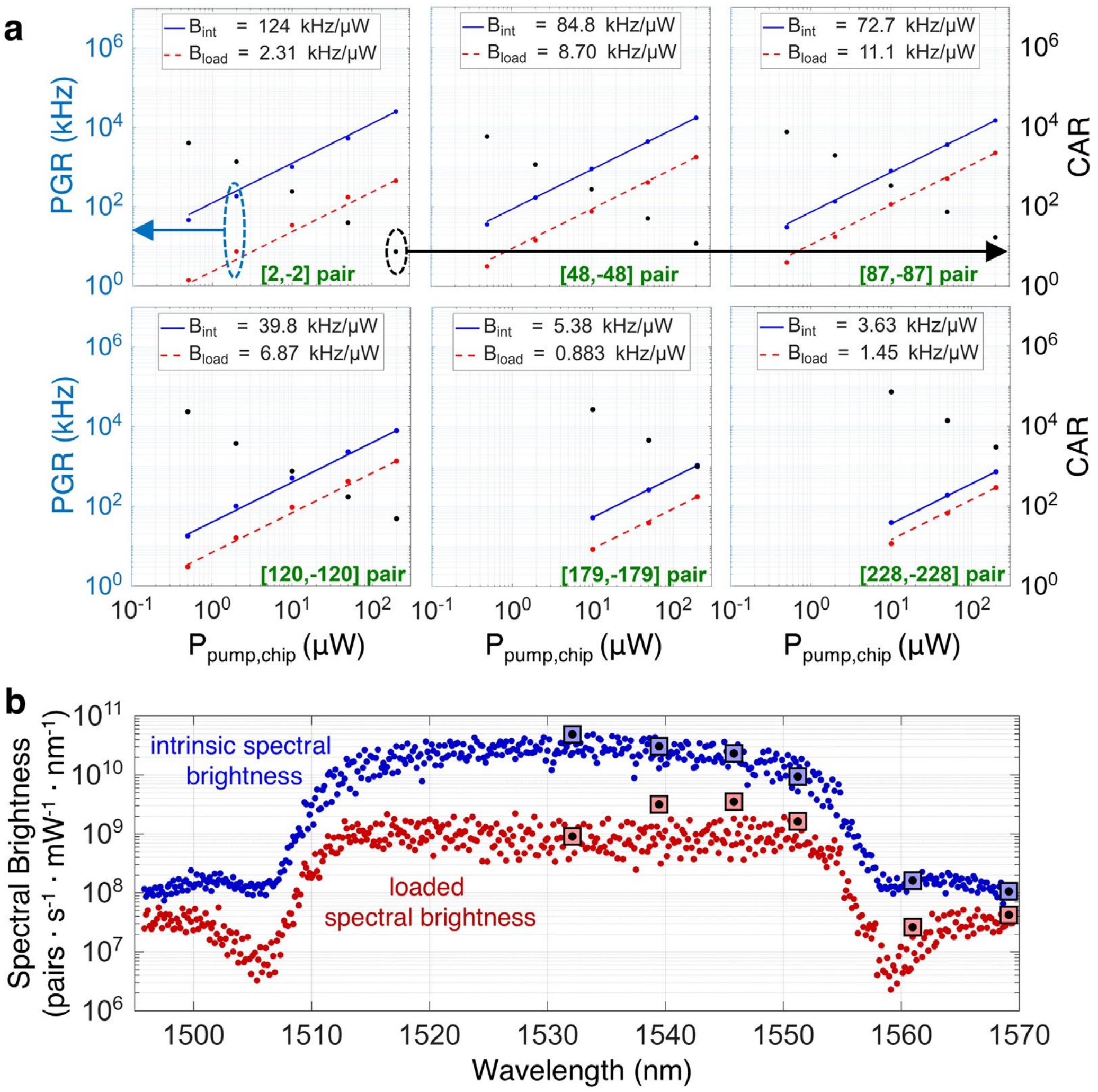

**a** Intrinsic PGR, loaded PGR, and CAR measured as a function of pump power for six representative conjugate frequency-bin pairs. Solid and dashed lines are linear fits to $PGR_{\text{int}}$ and $PGR_{\text{load}}$, respectively. **b** Spectral $B_{\text{int}}/\Delta\lambda$ (blue dots) and $B_{\text{load}}/\Delta\lambda$ (red dots) across the SPDC quantum frequency comb. Circles indicate values estimated from mode-resolved PGRs at a pump power of 200 μW, and squares denote the spectral brightnesses obtained from the linear fits in panel **a**.

To further assess the single-photon character of the frequency-bin-resolved heralded output, we measured the heralded second-order autocorrelation, $g_H^{(2)}(0)$, for three representative signal–idler bin pairs: [2,−2], [120, −120], and [228, −228], sampling the undercoupled, near-critical, and overcoupled regimes. Figure 5a shows the pump-power dependence of $g_H^{(2)}(0)$ for these bin pairs. The increase of $g_H^{(2)}(0)$ with pump power is consistent with a larger probability of additional pair generation and accidental coincidences within the measurement window. At a pump power of 89.3 μW, the undercoupled pair [2,−2] exhibits $g_H^{(2)}(0) = 0.087 \pm 0.016$; for the near-critically coupled pair [120, −120], $g_H^{(2)}(0) = 0.024 \pm 0.0036$; and for the overcoupled pair [228, −228], $g_H^{(2)}(0) = 0.0004 \pm 0.00011$. The corresponding $PGR_{\mathrm{int}}$ values are 8.62 MHz, 4.66 MHz, and 332 kHz, respectively.

**Fig. 5: Heralded second-order autocorrelation**

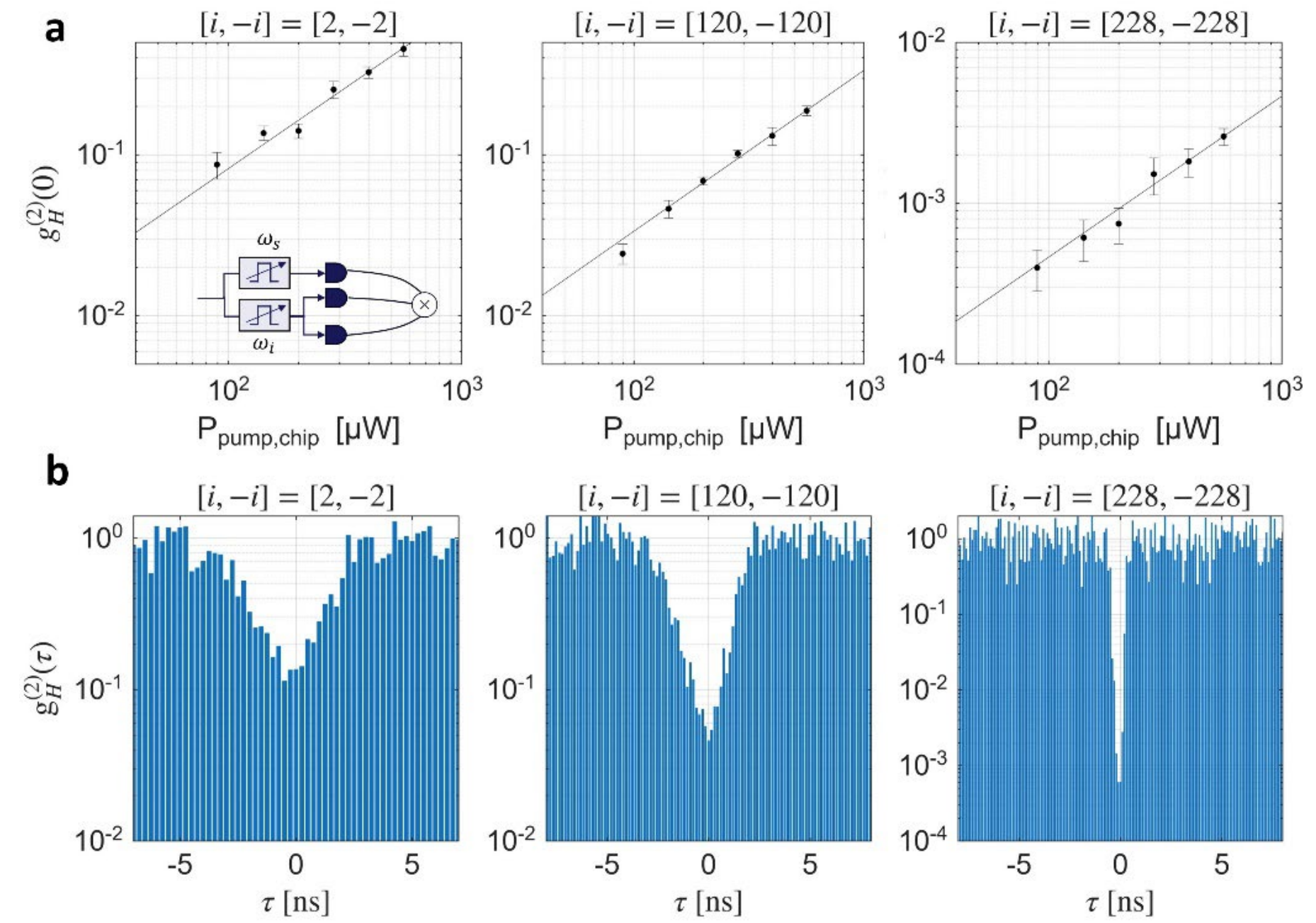

**a** Heralded second-order autocorrelation, $g_H^{(2)}(0)$, as a function of on-chip pump power for the selected bin pairs [2, −2], [120, −120], and [228, −228], representing the undercoupled, near-critically coupled, and overcoupled loading regimes, respectively. **b** Representative delay-dependent heralded second-order autocorrelation histograms, $g_H^{(2)}(\tau)$, measured at $P_{\mathrm{pump}}$ = 142 μW for the same selected bin pairs.

Figure 5b shows representative delay-dependent heralded autocorrelation histograms, $g_H^{(2)}(\tau)$, measured at a pump power of 142 μW for the same three selected bin pairs. Moving from the undercoupled to the near-critical and overcoupled loading regimes, the histograms exhibit a progressively narrower temporal feature and a deeper suppression around $\tau = 0$. This trend is consistent with the corresponding variations of the signal-idler coherence time and CAR observed for the selected bins in Fig. 3c.

## Cavity-loading control of the loaded PGR-CAR relation

We examine how the distinction between $PGR_{\text{int}}$ and $PGR_{\text{load}}$, corresponding to $B_{\text{int}}$ and $B_{\text{load}}$, affects the CAR of the generated photon pairs through the joint extraction efficiency $\eta_{\text{joint}}^{(i)}$ of each conjugate frequency-bin pair. Figure 6a shows the CAR as a function of $PGR_{\text{int}}$ for frequency-bin pairs across the comb. The conventional intrinsic-PGR–CAR trade-off is clearly observed: the CAR decreases as $PGR_{\text{int}}$ increases. For a fixed coincidence window, at high $PGR_{\text{int}}$, the probability of generating more than one photon pair within the coincidence window increases, leading to larger accidental coincidences and reduced CAR. Conversely, lower $PGR_{\text{int}}$ suppresses multi-pair events and yields higher CAR. This plot therefore captures the intrinsic multi-pair-limited behavior of the cavity-enhanced SPDC source.

However, a different relation appears when the CAR is plotted against $PGR_{\text{load}}$, as shown in Fig. 6b. $PGR_{\text{load}}$ includes the joint extraction efficiency $\eta_{\text{joint}}$ and therefore does not have to follow the same monotonic trend as $PGR_{\text{int}}$. In the spectral groups highlighted by the blue and red dashed lines, the CAR increases with increasing $PGR_{\text{load}}$, whereas the intermediate region shows the opposite trend. This behavior can be understood by separating pair-generation statistics from photon extraction. For a fixed coincidence window, the CAR is primarily set by $PGR_{\text{int}}$, because true coincidences scale as $PGR_{\text{int}}\eta_{\text{joint}}$, whereas multi-pair accidentals scale as $(PGR_{\text{int}})^2\eta_{\text{joint}}$. Thus, increasing $\eta_{\text{joint}}$ at a similar $PGR_{\text{int}}$ increases $PGR_{\text{load}}$ while maintaining nearly the same CAR, whereas reaching the same $PGR_{\text{load}}$ by simply increasing $PGR_{\text{int}}$ would reduce CAR through multi-pair noise.

Figures 6c and 6d provide the comb-resolved view of the two quantities underlying this separation: $PGR_{\text{int}}$ and $\eta_{\text{joint}}$. Near the center of the photonic bandgap (the blue dashed guide region), the undercoupled high-Q modes provide large $PGR_{\text{int}}$ but small $\eta_{\text{joint}}$. Toward and beyond the bandgap edge (the red dashed guide region), the resonances become more strongly loaded, increasing $\eta_{\text{joint}}$, while $PGR_{\text{int}}$ decreases because of broader resonance linewidths. Within each dashed guide region in Fig. 6b, $PGR_{\text{int}}$ remains within a similar range while $\eta_{\text{joint}}$ increases (Fig. 6d), shifting the data points toward larger $PGR_{\text{load}}$ without a proportional increase in multi-pair noise. In the intermediate region, the reduction in $PGR_{\text{int}}$ dominates over the increase in extraction efficiency, causing the CAR to increase even as $PGR_{\text{load}}$ decreases.

**Fig. 6: Cavity-loading dependence of the CAR-PGR relation**

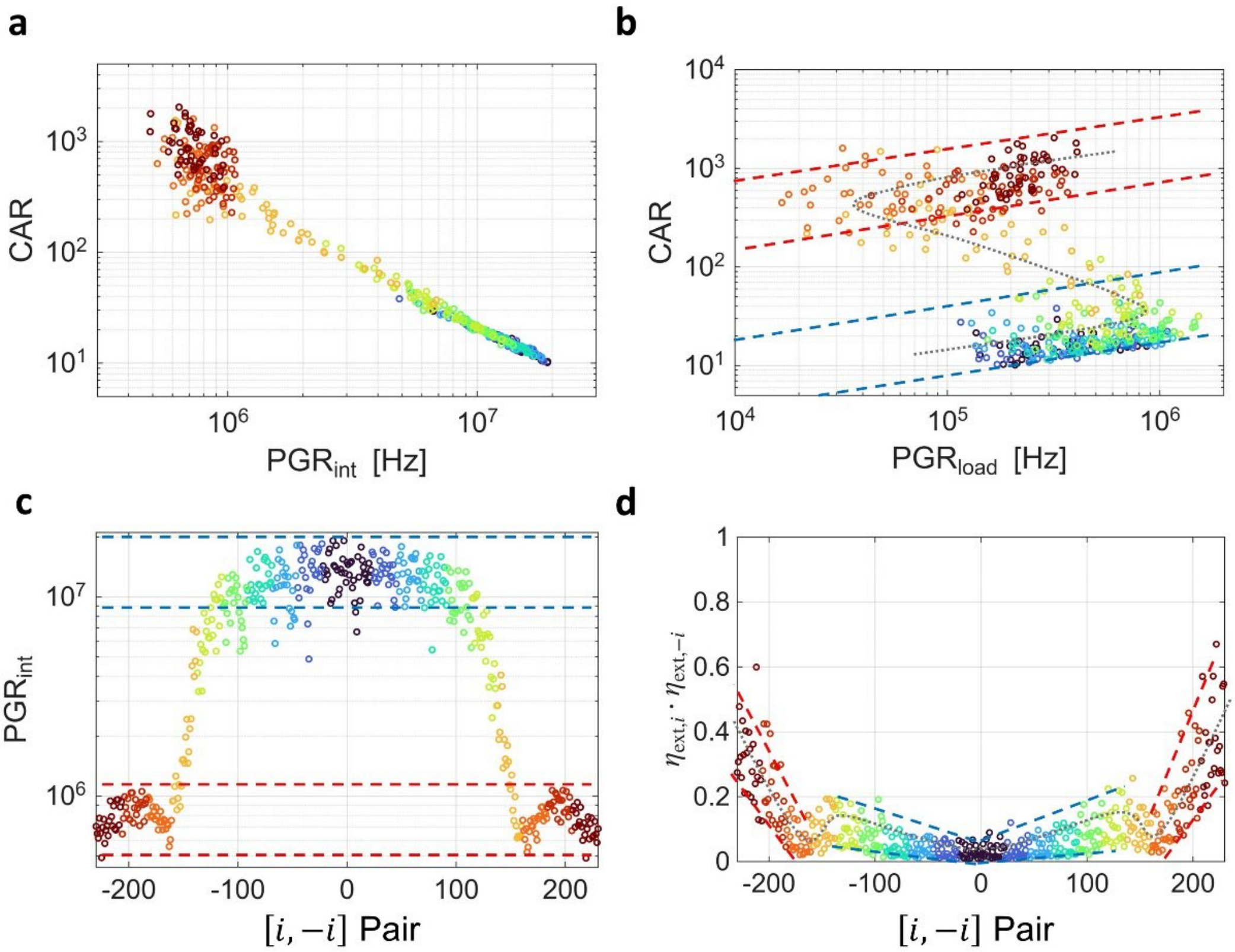

**a** Coincidence-to-accidental ratio (CAR) as a function of intrinsic photon-pair generation rate ($PGR_{\text{int}}$), for conjugate frequency-bin pairs $[i, -i]$. **b** CAR as a function of loaded photon-pair generation rate ($PGR_{\text{load}}$). Dashed guide regions highlight different cavity-loading regimes. **c** Comb-resolved $PGR_{\text{int}}$ as a function of the channel-1 index $i$. **d** Joint extraction efficiency, $\eta_{\text{joint}} = \eta_{\text{ext},i} \cdot \eta_{\text{ext},-i}$. The on-chip pump power is fixed to 200 μW.

This analysis shows why $PGR_{\text{load}}$, rather than $PGR_{\text{int}}$ alone, is the relevant metric for designing usable cavity-enhanced SPDC QFCs. Maximizing intrinsic generation with the highest-Q undercoupled modes can produce large internal pair rates, but only a small fraction of the generated photons is extracted.

Stronger output loading improves the usable photon-pair flux, whereas excessive loading reduces the resonant enhancement and limits the absolute generation rate. It is therefore expected that, depending on the target application, relevant parameters such as resonant enhancement, extraction efficiency, and multi-pair noise level should be properly customized. The present device maps their behavior across under-, critical-, and over-coupled cavity-loading regimes. Engineered PhC mirrors can co-locate high $PGR_{\text{int}}$ and high $\eta_{\text{joint}}$ within the photonic bandgap, enabling brighter high-CAR SPDC QFCs.

## Pump-selectable frequency-bin pairing

Beyond the generation of a dense QFC, the single-pass pump configuration of the PhC-FP microresonator provides a practical advantage to reconfigure the correlated frequency-bin pairs. In triply resonant schemes, the pump, signal, and idler fields must be simultaneously matched to cavity resonances. Because the short-wavelength pump and telecom signal-idler modes experience different dispersion, retuning the pump to select a different bin-pair family generally cannot preserve simultaneous resonance; the unavoidable pump-resonance constraint strongly modulates the overall QFC operation. In our PhC–FP configuration, by contrast, the visible pump propagates in a single pass and is not constrained by a cavity resonance. The pump frequency therefore acts as an independent control parameter: different resonant signal–idler mode combinations can be selected through energy conservation while the telecom-band cavity resonance grid remains fixed.

Figure 7a shows the working principle of pump-selectable frequency-bin pairing. For a reference pump frequency $\omega_{p,0}$, the generated QFC is centered at $\omega_{p,0}/2$, and the correlated bin pairs are denoted by $[i, -i]$. We use the same frequency-bin index $i$ as in Fig. 3, with $i$ increasing toward longer wavelength (decreasing toward higher frequency). When the pump frequency is tuned to $\omega_p = \omega_{p,0} + m\Delta\omega_{FSR}$, the energy conservation relation $\omega_i + \omega_j = \omega_p$ gives $i + j = -m$. Here, the integer $m$ denotes the pump-selected frequency-bin pairing condition and defines the correlated frequency-bin pairing as $[i, -i - m]$. Pump detuning therefore translates the zero-detuning pairing $[i, -i]$ across the fixed cavity resonance grid.

**Fig. 7: Pump-selectable frequency-bin pairing**

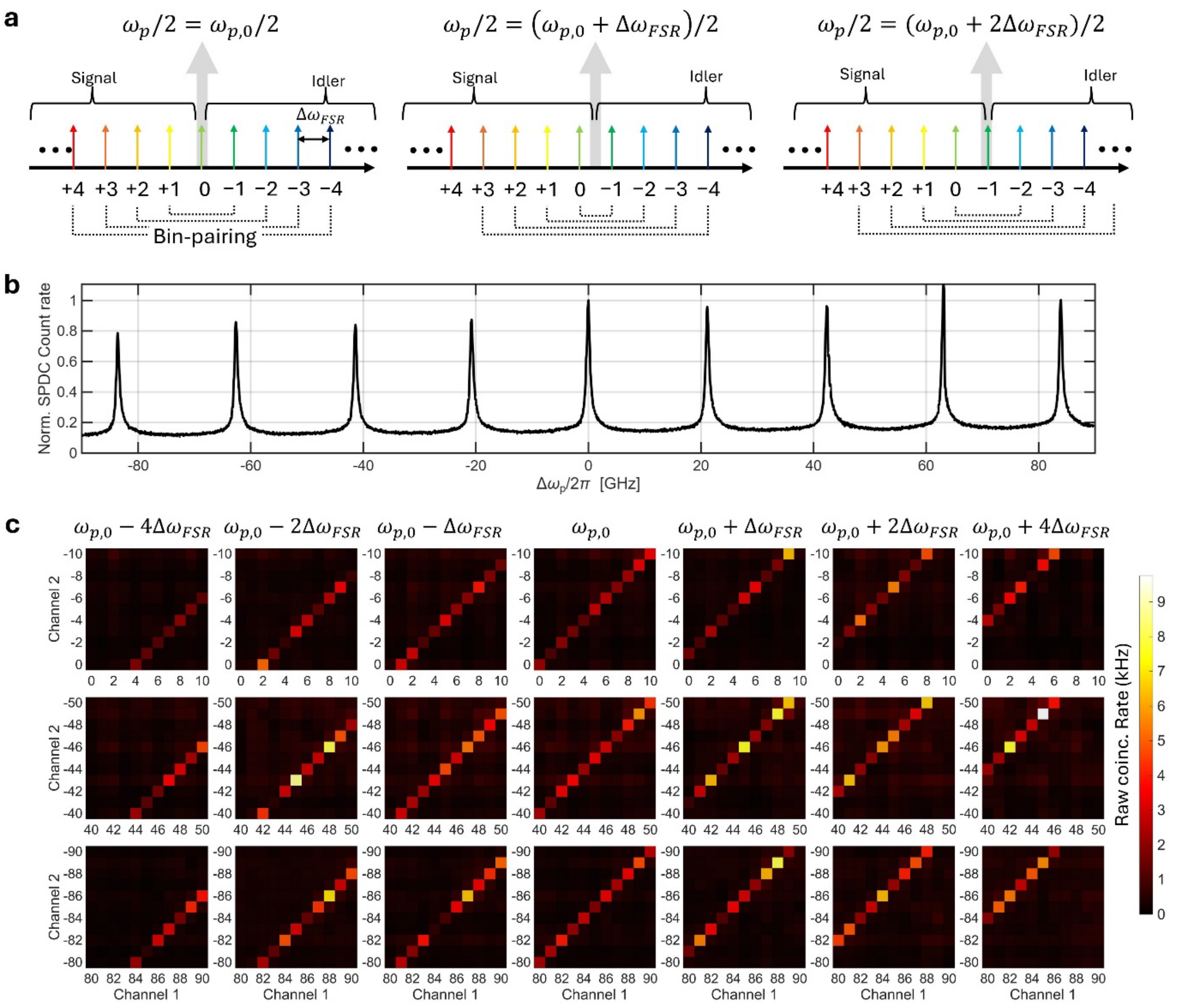

**a** Schematic of pump-selectable frequency-bin pairing in the fixed signal-idler resonance ladder. When the pump frequency is detuned by an integer multiple of the PhC-FP resonant mode spacing, $\Delta\omega_{\mathrm{FSR}}$, the correlation diagonal of the quantum frequency comb translates according to $i + j = -m$, where $m$ is the pump-selected pairing condition. **b** Normalized mode-integrated, single-channel SPDC signal count rate as a function of pump-frequency detuning, $\Delta\omega_p/2\pi$. **c** Joint spectral intensity matrices measured at seven selected pump-detuning conditions, $m = -4, -2, -1, 0, +1, +2$, and $+4$, in three selected windows indexed by the channel-1 ranges, $i = 0 - 10$, $40 - 50$, and $80 - 90$. The on-chip pump power is 200 μW.

To experimentally verify this pump-controlled reconfiguration, we measured the mode-integrated single-channel SPDC signal as a function of pump-frequency detuning, $\Delta\omega_p/2\pi$, as shown in Fig. 7b. The count rate was normalized to the value obtained at the reference pump wavelength of 766 nm, corresponding to the zero-detuning pairing condition $m = 0$. The measured response shows nine periodic peaks from $m = -4$ to $m = +4$, with a spacing of approximately 21 GHz, in good agreement with the FSR of the signal-idler resonance ladder. These peaks indicate pump settings for which a family of cavity-defined signal-idler bin pairs satisfies energy conservation. The measured full-width at half maximum of approximately 769 MHz is consistent with the effective convolution of the participating signal and idler resonance linewidths. Notably, the nine pump-selected peaks exhibit comparable mode-integrated SPDC count rates, demonstrating practical bin-pair reconfiguration while preserving QFC brightness.

To further investigate the reconfiguration of correlated frequency-bin pairs, we measured the joint spectral intensity matrices for three 11 × 11 bin-pair windows indexed by the channel-1 ranges ($i = 0 - 10$, $40 - 50$, and $80 - 90$), under seven representative pump-detuning conditions: $m = -4, -2, -1, 0, +1, +2,$ and $+4$, as shown in Fig. 7c. As the pump frequency is detuned, the correlation diagonal in the joint spectral intensity matrix translates systematically according to $i + j = -m$, which confirms that the correlated bin-pair family is selected by the pump while the cavity resonance grid remains fixed. The pump-detuning experiment shows that the single-pass pump not only drives photon-pair generation, but also serves as a control knob for programming the frequency-bin correlation structure. Combining pump shaping[60] with cavity-loading control provides a route toward tailored QFCs with programmable pairing rules, pair-dependent weights, and clustered frequency-bin correlations.

## Conclusion

We demonstrated a high-brightness PPLN PhC-FP microresonator platform for single-pass-pumped, resonator-enhanced SPDC QFC generation. The platform combines single-pass visible pumping with resonant enhancement of telecom-band signal and idler photons, enabling efficient photon-pair generation while relaxing simultaneous pump-signal-idler spectral-alignment constraints and simplifying scalable integration of frequency-multiplexed sources. Reciprocal SHG measurements verify the quasi-phase-matched nonlinear gain and its overlap with the cavity resonances, while mode-integrated SPDC measurements yield an intrinsic brightness of 69.9 MHz/μW and a loaded brightness of 1.88 MHz/μW, together with a maximum CAR of 16,000 at $P_{\mathrm{pump}}$ = 1.23 nW. Frequency-resolved measurements reveal a dense QFC spanning 1495-1570 nm with 461 cavity-defined frequency bins, strong pairwise frequency correlations, and intrinsic and loaded spectral brightness values exceeding ~$4.37\times10^{10}$ and ~$2.22\times10^{9}$ pairs/(s·mW·nm), respectively. In a selected high-CAR cavity-defined bin pair, we measure a heralded second-order autocorrelation of $g_H^{(2)}(0) = (4.0 \pm 1.1) \times 10^{-4}$.

Beyond these performance metrics, the PhC-FP platform provides two complementary design handles for QFC sources. First, the mode-resolved analysis shows that $PGR_{\text{int}}$ and $PGR_{\text{load}}$ are governed by different aspects of the cavity response; high-Q undercoupled modes maximize internal generation, whereas stronger loading improves photon extraction. The resulting extraction-aware $PGR_{\text{load}}$–CAR analysis identifies $PGR_{\text{load}}$, rather than $PGR_{\text{int}}$ alone, as the relevant metric for designing practical cavity-enhanced SPDC QFCs. By exploiting a fundamental-mode-selective resonance ladder and the tailorable reflection response of PhC mirrors, the PhC-FP geometry can position the target loading regime within the photonic bandgap and engineer photon-pair properties such as the Schmidt number and spectral purity. Second, the non-resonant single-pass pump selects the correlated frequency-bin pairing on a signal-idler grid deterministically defined by the PhC–FP resonant modes. Spectral and temporal pump shaping can extend this single-diagonal pairing to preprogram selected correlation-matrix diagonals and pair-dependent weights at generation. Combined with extraction-aware cavity-loading control, such tailored SPDC QFCs can provide chip-integrated non-classical light resources that downstream photonic circuits can utilize to implement for multichannel-interference quantum simulation[10,11,13,21,22,61], programmable frequency-bin information processing[29,50,62,63,64], and field-quadrature-based quantum-state engineering.[65,66,67,68,69]

# Methods

## Sample Fabrication

The device was fabricated on a 500-nm-thick 5 mol% MgO-doped lithium niobate-on-insulator wafer supplied by NANOLN. Periodic poling was first performed by applying high-voltage electrical pulses through patterned Cr/Au electrodes (10 nm/200 nm) deposited on the bare LNOI surface by thermal evaporation. After verifying the poling quality across all electrode regions, the Cr/Au electrodes were removed using gold and chromium etchants, while the alignment markers were preserved for subsequent lithography steps.

The photonic structures were then defined by electron-beam lithography using a FOX-16 negative-tone resist mask aligned to the pre-defined metal alignment markers. Pattern transfer was carried out using $Ar^+$ inductively coupled plasma reactive-ion etching (ICP-RIE) to form the lithium niobate waveguides and photonic-crystal structures. Following etching, the residual FOX-16 mask was removed using buffered oxide etchant, and the sample was subsequently cleaned in Standard Clean-2 solution to remove etch redeposition and surface contaminants.

Finally, a shallow etching process using Ar-ion and $O_2$ plasma was performed to further smooth the device sidewalls and precisely adjust the waveguide thickness for quasi-phase matching. The resulting devices consist of periodically poled lithium niobate waveguides integrated with photonic-crystal Fabry–Pérot cavities for enhanced nonlinear frequency conversion and photon-pair generation.

## Measurements

SPDC photon-pair generation was driven by a tunable pump laser (Newport, TLB-6712) operating at 765–781 nm. A 1% pick-off was monitored by a Fizeau wavemeter (MogLabs, FZW600) and used for PID wavelength stabilization, while the main beam passed through a variable optical attenuator and polarization controller before lensed-fiber coupling into the input facet; a second 1% pick-off monitored the launched pump power. The device was held at 60 °C. Telecom-band SPDC photons were collected from the output facet using a lensed fiber (2.5 μm spot diameter), separated from residual pump light by a fiber-based wavelength demultiplexer (HaphiT, FCLD-7815-1-L10-FC/APC) and three cascaded long-pass filters (Thorlabs, FELH1050), and detected by superconducting nanowire single-photon detectors (Quantum Opus) after a 50:50 beam splitter. Polarization controllers in the output arms optimized the detector coupling. Single-photon counts and coincidences were recorded by a fast time tagger (Swabian Instruments, Time Tagger Ultra). Frequency-bin-resolved measurements used two tunable band-pass filters (WL Photonics Inc., WLTF-NE-S-1550) in the two output arms to select the signal and idler bins before detection.

Telecom-band reflectance and reciprocal SHG measurements used a tunable laser (Toptica, CTL1550, 1510–1620 nm) injected into the output-coupler facet, corresponding to the SPDC output port, through a variable optical attenuator, polarization controller, fiber circulator and lensed fiber. A 1% pick-off was sent to a fiber Mach–Zehnder interferometer with a 1 m path-length imbalance to calibrate wavelength shifts during laser scans; another 1% pick-off before the circulator monitored the launched telecom laser power. The circulator routed the reflected telecom signal to a photodetector (Thorlabs, DET10C/M) for extracting infrared cavity linewidths. For reciprocal SHG, the generated second-harmonic light was collected from the opposite facet, corresponding to the SPDC-pump input port, using the same lensed fiber. The second-harmonic was separated from the residual fundamental field with the same fiber-based wavelength demultiplexer and measured using a femtowatt photoreceiver (Newport 2151).

## Data Availability

The data that support the findings of this study are available from the corresponding author upon reasonable request.